\documentclass[11pt,document,nofootinbib,superscriptaddress,onecolumn,preprintnumbers,balancelastpage]{article}

\pdfoutput=1
\usepackage{jhep}
\usepackage{cancel}
\usepackage[normalem]{ulem}
\usepackage{slashed}
\usepackage{mathtools}
\usepackage{amsfonts}
\usepackage{amssymb}
\usepackage{amsmath}
\usepackage{scalerel}

\usepackage{epsfig}
\usepackage{rotating}
\usepackage{url}
\usepackage{times}
\usepackage{color}
\usepackage{bm}
\usepackage{xcolor,colortbl}
\usepackage{hyperref}
\usepackage{enumitem}
\usepackage{fancyhdr}
\usepackage{anyfontsize}
\usepackage{wasysym}
\usepackage{empheq}
\usepackage{comment}

\newcommand{\EM}
{\scalerel*{M}{t}}
\newcommand{\AR}
{\scalerel*{R}{t}}
\newcommand{\one}{\scalerel*{1}{Y}}
\newcommand{\two}
{\scalerel*{2}{t}}

\newcommand{\be}{\begin{equation}}
\newcommand{\ee}{\end{equation}}
\newcommand{\bea}{\begin{eqnarray}}
\newcommand{\eea}{\end{eqnarray}}

\newcommand{\lb}{\boldsymbol{\ell}}
\newcommand{\nb}{\mathbf{n}}
\newcommand{\mb}{\mathbf{m}}
\newcommand{\eb}{\mathbf{e}}

\def\tbm{{\bf \tilde{m}}}

\let\isout\sout \renewcommand{\sout}[1]{\ifmmode\text{\isout{\ensuremath{#1}}}\else\isout{#1}\fi}

\definecolor{colorTM}{rgb}{.2,.7,.2}

\newcommand{\kibitz}[2]

\title{Taub-NUT Instanton as the Self-dual Analog of Kerr}

\author[a]{Jash Desai,}
\author[a]{Gabriel Herczeg,}
\author[b]{David McNutt,}
\author[a]{and Max Pezzelle}

\affiliation[a]{\footnotesize Brown Theoretical Physics Center, Department of Physics, Brown University, Providence, RI 02912, USA}
\affiliation[b]{\footnotesize Center for Theoretical Physics, Polish Academy of Sciences, Warsaw}

\date{\today}
\abstract{It was recently conjectured that a certain vacuum Kerr-Schild spacetime, which may be regarded as a self-dual analog of the Kerr metric, is equivalent to the self-dual Taub-NUT instanton. We confirm this conjecture by applying the Cartan-Karlhede algorithm to each metric and showing that for suitable choices of null tetrad, the algorithm leads to the same invariants and linear isotropy groups for both, establishing their equivalence. While it is well-known that the Taub-NUT solution and its self-dual version admit a \emph{double} Kerr-Schild form, the observation that the self-dual Taub-NUT instanton admits a single Kerr-Schild form has only been made very recently. The two metrics we compare may be regarded as either complex metrics with Lorentzian $(1,3)$ signature or real metrics with Kleinian $(2,2)$ signature; here we take the latter view. Significant simplifications occur when the null tetrads are chosen to consist of two pairs of complex conjugate null vectors rather than four real independent ones. As a bonus, our work provides the first example of applying the Cartan-Karlhede algorithm using a null tetrad of this type.}

\begin{document}

\maketitle

\section{Introduction}
The historical significance of the Kerr-Schild ansatz for the discovery of some of the most important exact solutions of Einstein's equations is hard to overstate---most notably, the Kerr-Schild form was crucial to the discovery of the Kerr metric and higher dimensional rotating black hole solutions of Myers and Perry \cite{PhysRevLett.11.237, Kerr:2007dk,  myers1986black}. Recently, interest in Kerr-Schild metrics has been reignited by the color-kinematics duality approach to relating scattering amplitudes in gauge theories and gravity \cite{Bern:2008qj}, and its analog for classical exact solutions,  the classical double copy \cite{Monteiro:2014cda, Luna:2018dpt, White:2020sfn, Chacon:2021wbr, Elor:2020nqe, Farnsworth:2021wvs, Farnsworth:2023mff, Easson:2023dbk, Campiglia:2021srh, Adamo:2021dfg, Godazgar:2021iae, Mao:2023yle, Luna:2015paa, Chawla:2022ogv, Easson:2020esh, Easson:2021asd, Ilderton:2024oly}.

Analytic continuation from Lorentzian $(1,3)$ signature to Euclidean $(0,4)$ is by now a standard technique in thermodynamics and quantum field theory. However, in the last few years it has become apparent that analytic continuation to Kleinian $(2,2)$ signature\footnote{In the mathematics literature, Kleinian spaces are often said to have \emph{neutral signature}, and in a slightly more general context, are referred to as \emph{neutral geometries} \cite{hervik2010pseudo,Brooks:2015lba}.}  is a natural approach to computing scattering amplitudes in four dimensions, where on-shell three-point amplitudes, which can only be non-trivial in Kleinian signature, are useful for recursively constructing higher-point on-shell amplitudes \cite{Britto:2005fq, Atanasov_2021}. For this reason, and others, classical solutions of Einstein's equations with Kleinian signature have attracted significant interest in recent years.

One of the features that distinguishes Kleinian signature solutions from their Lorentzian counterparts is that there are no real, self-dual solutions in Lorentzian signature apart from those that are conformally flat. By contrast, in Kleinian signature, the left and right Weyl spinors are real and independent (at least if the null tetrad is chosen to be real), and non-trivial Kleinian instantons are possible. One such instanton is the self-dual Taub-NUT solution, which was recently shown by Crawley, Guevara, Miller and  Strominger to be diffeomorphic to the self-dual Kerr-Taub-NUT metric, rendering the angular momentum of black holes pure gauge in the self-dual sector \cite{Crawley:2021auj, Crawley:2023brz}. In \cite{Easson:2023ytf}, a Kleinian signature analog of the Birkhoff theorem was established.

In \cite{Easson:2023dbk}, the role of isometries in the classical double copy was clarified by reformulating the Kerr-Schild double copy in the language of the Newman-Penrose formalism. One of the benefits of this reformulation is that it provides a well-defined mapping between real vacuum Kerr-Schild metrics and self-dual ones. An immediate application of this map between real and self-dual solutions is the ability to write down a self-dual analog of the Kerr metric. The resulting metric is complex in Lorentzian signature, but it can be analytically continued to a real metric in Kleinian signature. This example was studied in \cite{Easson:2023dbk}, where it was noted that the angular momentum can be gauged away with a simple diffeomorphism to produce a self-dual solution depending only on a single real parameter, at which point it was conjectured that the resulting solution is equivalent to the Kleinian self-dual Taub-NUT metric. 

The main purpose of this article is to prove that the self-dual analog of Kerr is indeed equivalent to the self-dual Taub-NUT instanton, a result which affirms a very recent paper showing that self-dual Taub-NUT admits a single Kerr-Schild form \cite{Kim:2024dxo}. We establish this equivalence by making use of the Cartan-Karlhede algorithm, which, roughly speaking, classifies spacetimes according to certain invariant quantities associated with a suitable choice of null or orthonormal tetrad. If all of the invariants generated from the algorithm coincide, the spacetimes are equivalent. When two metrics are equivalent, there always exists a diffeomorphism taking one to the other. However, as a practical matter, it is often extremely difficult to determine an explicit diffeomorphism between them. In comparison, the Cartan-Karlhede algorithm provides a straightforward procedure to determine when two metrics are equivalent without having to find a transformation connecting them.

The paper is organized as follows: In section \ref{sec:CK Algorithm} we give an overview of the Cartan-Karlhede algorithm, paying special attention to Kleinian signature solutions with null tetrads consisting of two pairs of complex conjugate (dual) vectors. In section \ref{sec:review}, we review the self-dual analog of Kerr. In section \ref{sec:applyCK}, we apply the algorithm  to characterize the self-dual analog of Kerr and the self-dual Taub-NUT metrics. In section \ref{sec:conclusions} we apply the results of the previous section to establish equivalence of the two solutions and conclude with a brief discussion of our results.

\section{Characterizing Kleinian spacetimes} \label{sec:CK Algorithm}

To compare two Kleinian spacetimes, we must characterize the geometry of each and compare their respective invariants. For a given Kleinian geometry, one can choose an orthonormal basis $\{\eb^a\}$ such that
\be
	ds^2 = (\eb^1)^2 + (\eb^2)^2 - (\eb^3)^2 - (\eb^4)^2, 
\ee
or equivalently, a null basis $\{\lb^1, \lb^2, \nb^1, \nb^2\}$, in terms of which the metric takes the form
\be
    ds^2 = 2\lb^1 \nb^1 - 2 \lb^2 \nb^2. \label{4dneutral} 
\ee
The null basis elements can be expressed in terms of the orthonormal ones as
\bea 
\lb^1 &=& \frac{1}{\sqrt{2}}(\eb^1 + \eb^3), \quad \nb^1 
 = \frac{1}{\sqrt{2}}(\eb^1 - \eb^3) \\
 \lb^2 &=& \frac{1}{\sqrt{2}}(\eb^2 + \eb^4), \quad \nb^2 
 = \frac{1}{\sqrt{2}}(\eb^2 - \eb^4).
\eea 
\noindent Relative to the null basis, there is a six parameter group of $SO^{*}(2,2)$ transformation that leave the metric invariant \cite{Brooks:2015lba}, which can be divided into null rotations
\be 
\begin{aligned}
	& {\lb^1}' = \lb^1 + B_1 \lb^2 + B_2 \nb^2 + B_1 B_2 \nb^1, 			\quad {\nb^1}' = \nb^1, \\
	& {\lb^2}' = \lb^2 + B_2 \nb^1, \quad {\nb^2}' = \nb^2 	+ B_1 	\nb^1, \\
	& \\
	& {\lb^1}' = \lb^1, \quad {\nb^1}' = \nb^1 + C_1 \lb^2 	+ C_2 \nb^2 + C_1 C_2 \lb^1, \\
	& {\lb^2}' = \lb^2 + C_2 \lb^1, \quad {\nb^2}' = \nb^2 + 			C_1 	\lb^1,
\end{aligned} \label{NullRot}
\ee
\noindent and boosts
\be
\begin{aligned}
	& {\lb^1}' = A_1 \lb^1,\quad {\nb^1}' = A_1^{-1} \nb^1,\\
	&{\lb^2}' = A_2 \lb^2,\quad {\nb^2}' = A_2^{-1} \nb^2.
\end{aligned} \label{NullBst}
\ee
\noindent Here $A_1, A_2, B_1, B_2, C_1$ and $C_2$ are real-valued parameters for the frame transformations.

Usually, the basis elements are taken to be real-valued, but a complex-conjugate basis is also permitted
\be
    ds^2 = 2\mb^1 \bar{\mb}^1 - 2\mb^2 \bar{\mb}^2. \label{4dneutral_complex} 
\ee
\noindent This leads to a different representation of the $SO^{*}(2,2)$ transformation group. The corresponding transformations for the null rotations are quite complicated, but can be recovered from the real-valued null basis in equation \eqref{NullRot}. 
The boost transformations now resemble rotations in two complex planes:
\be
\begin{aligned}
	& \mb^{1'} = e^{iA_1} \mb^1,\quad \bar{\mb}^{1'} = e^{-iA_1} \bar{\mb}^1,\\
	&\mb^{2'} = e^{iA_2} \mb^2,\quad \bar{\mb}^{2'} = e^{-iA_2} \bar{\mb}^2.
\end{aligned} \label{NullBstC}
\ee

In this paper we will use tensor quantities to characterize the Kleinian spacetimes we wish to compare. In practice, it would be beneficial to switch to a spinor representation of the tensor quantities as this would reduce the number of components to be computed. A spinor approach to Kleinian geometry can be naturally adapted from the Newman-Penrose approach used in Lorentzian geometry by considering the double cover of the corresponding structure group. This has been developed in \cite{law2006classification, law2009spin}, which is related to the newer spinor formalism in  \cite{Crawley:2021auj}. 

Using a temporary abuse of notation, we will denote the real-valued basis and complex-valued basis with the same labels:
\be
    \begin{aligned}
        &\lb = \lb^1 , \qquad \nb = \nb^1  , \qquad \mb = \lb^2 , \qquad \tbm =  \nb^2, \\ &\text{\hspace{3.5cm} or} \\
        &\lb = \mb^1 , \qquad \nb = \bar{\mb}^1  , \qquad \mb = \mb^2 , \qquad \tbm =  \bar{\mb}^2.
    \end{aligned}
\ee
\noindent The left Weyl scalars are
\be
    \begin{aligned}
        \Psi_0 & = C_{abcd} \lb^a \mb^b \lb^c \mb^d \\
        \Psi_1 & = C_{abcd}\lb^a \nb^b \lb^c \mb^d \\
        \Psi_2 & = - C_{abcd} \lb^a \mb^b \nb^c \tbm^d  \\
        \Psi_3 & = C_{abcd}\nb^a \lb^b\nb^c \tbm^d \\
        \Psi_4 & = C_{abcd}\nb^a \tbm^b \nb^c \tbm^d, \\
    \end{aligned} \label{eq:WeylScalarTemplate}
\ee
\noindent while the right Weyl scalars can be obtained by swapping $\mb$ with $\tbm$ in the above expressions. 

The components of the covariant derivatives of the Weyl tensor may be expressed in terms of the spin-coefficients and the derivatives of the Weyl scalars. For the moment, we will only display the tensor components for higher covariant derivatives.

\subsection{The Cartan-Karlhede algorithm}

Any pseudo-Riemmannian space can be classified by a finite set of invariants. These invariants are known as \emph{Cartan invariants} and they are determined from the components of the curvature tensor and its covariant derivatives, relative to a common coframe. This coframe is determined in a coordinate independent manner by imposing certain geometric conditions.  Such a coframe is said to be \emph{invariantly defined}. The process of determining an invariantly defined coframe can be formalized into an algorithm, known as the Cartan-Karlhede algorithm. This algorithm has been implemented for the Lorentzian case \cite{Kramer} and the Kleinian case \cite{Brooks:2015lba}. 

To determine the geometric conditions on the coframe, the algorithm exploits the canonical forms of the irreducible parts of the curvature tensor at iteration 0, or the $q^{th}$ covariant derivatives of the curvature tensor at iteration $q$. At a given iteration, $q$, there is a subgroup of the frame transformation group that leaves the canonical forms of the curvature tensor and its covariant derivatives up to $q^{th}$ order unchanged, this subgroup is commonly known as the \emph{$q^{th}$ linear isotropy group}, $H_q$. In the Lorentzian case the alignment types give clear canonical forms; however, for more general pseudo-Riemannian spaces the larger non-compact group leads to a considerably larger selection of canonical forms using the alignment classification \cite{hervik2010pseudo}. In particular, the Petrov types of the neutral signature differ from the Lorentzian case \cite{law2006classification}. 

The algorithmic procedure for determining equivalence is as follows:

\begin{enumerate}
\item Set $q$, the order of differentiation, to 0.
\item Compute up to the the $q^{th}$ covariant derivatives of the Riemann tensor.
\item Fix the Riemann tensor and its covariant derivatives in a canonical form.
\item Fix the frame as much as possible using this canonical form, and record the {\it linear isotropy group $H_q$}. 
\item Find the number $t_q$ of independent functions in the components of the Riemann tensor and its covariant derivatives, in the canonical form.
\item If the number of independent functions, and the linear isotropy group  are the same as in the previous step, let $p+1=q$, and the algorithm terminates; if they differ (or if $q=0$), increase $q$ by 1 and go to step 2. 
\end{enumerate}

\noindent The components of the Riemann tensor and its covariant derivatives relative to the frame fixed at the end of the Cartan algorithm are called {\it Cartan invariants}. 

Thus, any four-dimensional (4D) spacetime is characterized by the discrete sequence of linear isotropy groups dimensions, and the discrete sequence of number of functionally independent invariants, the canonical forms used, and the Cartan invariants themselves. As there are $t_p$ essential spacetime coordinates, the remaining $4-t_p$ are ignorable. Then denoting the dimension of the linear isotropy group of the space-time as $s=\dim(H_p)$, the isometry group has dimension 
\be
	r=s+4-t_p. \label{IsoDim}
\ee

If two spacetimes have identical sets of Cartan invariants, the spacetimess are related by a coordinate transformation and we will say they are {\it equivalent}. To compare two spacetimes one can first compare the discrete sequences. If the sequences match for each metric, then the forms of the Cartan invariants relative to the same frame must be compared to determine equivalence. If one spacetime has a functionally dependent Cartan invariant, expressed in terms of the functionally independent Cartan invariants then the corresponding Cartan invariant for the other spacetime must share the same functional form in terms of the corresponding functionally independent Cartan invariants. Otherwise, the spacetimes are not equivalent. Cartan invariants that are functionally dependent on other Cartan invariants are known as \emph{classifying functions} for the geometry. 

\section{The Self-dual analog of Kerr} \label{sec:review}
Here we briefly review the properties of four-dimensional vacuum Kerr-Schild spacetimes in the Newman-Penrose formalism \cite{mcintosh1988single, mcintosh1989kerr, Kramer}, recall the map from real to self-dual Kerr-Schild solutions proposed in \cite{Easson:2023dbk}, and use this map to motivate the self-dual analog of Kerr.

A Kerr-Schild metric is one that can be written in form 
\be 
g_{ab} = \eta_{ab} + V\ell_a\ell_b \label{Kerr-Schild}
\ee 
where $V$ is a scalar function and $\ell_a$ is null with respect to both the full metric $g_{ab}$ and the background flat metric $\eta_{ab}$. For any Kerr-Schild metric, we can choose a null tetrad of the form
\be 
\begin{aligned}
	& \lb = dv + V\nb,  \\
	& \nb = du + Y d\tilde{\zeta} + \tilde{Y}d\zeta + Y\tilde{Y}dv, \\
	& \mathbf{m} = d \tilde{\zeta} + \tilde{Y}dv,\\
	& \tbm = d\zeta + Y dv , 
\end{aligned}
\ee
where $Y$ and $\tilde{Y}$ are scalar functions that are complex conjugates when the spacetime is real and Lorentzian, and real and independent when the spactime is real and Kleinian. Here we use totally null coordinates $(u,v,\zeta,\tilde{\zeta})$ which are related to the usual Cartesian coordinates by 
\be
\label{eq:coor}
u = \tfrac{1}{\sqrt{2}}(t - z), \quad v = \tfrac{1}{\sqrt{2}}(t+ z), \quad \zeta = \tfrac{1}{\sqrt{2}}(x + iy), \quad \tilde{\zeta} = \tfrac{1}{\sqrt{2}}(x - iy)
\ee
\noindent when the spacetime is Lorentzian, with similar expressions corresponding to the replacement ${y \to -iy}$ when the spacetime is Kleinian.

Now the metric
\be
g_{ab} = 2\ell_{(a}n_{b)}  - 2m_{(a}\tilde{m}_{b)} 
\ee
has the Kerr-Schild form \eqref{Kerr-Schild}, with background metric
\be 
ds_0^2 = \eta_{ab}dx^a dx^b = 2(du dv - d\zeta d\tilde\zeta).
\ee 
The conditions for such a Kerr-Schild metric to satisfy the vacuum Einsein equations are, firstly that $Y$ is the solution of the algebraic equation 
\cite{debney1969solutions} \be \label{fvac}
f(Y, u + Y\tilde{\zeta}, \zeta + Yv) = \Phi(Y) + (\tilde{c}Y + a)(\zeta + Yv) - (bY + c)(u + Y\tilde{\zeta}) = 0
\ee
where $\Phi(Y)$ is an arbitrary function, and $a,b,c,\tilde{c}$ are constants, with $c$ and $\tilde{c}$ being complex conjugates when the spacetime is real and Lorentzian, and independent and real when the spacetime is real and Kleinian. Second, the function $V$ must satisfy
\be  
V = \frac{m}{2P^3}(\rho + \tilde\rho), \label{Vvac}
\ee
where 
\be 
P = a + \tilde{c}Y + c\tilde{Y} + bY\tilde{Y},
\ee 
and $\rho$, $\tilde{\rho}$ are spin coefficients encoding the left and right complex expansions of $\ell$. Importantly, the spin coefficients $\rho$, $\tilde{\rho}$ are determined by the functions $Y$, $\tilde{Y}$ alone, so knowledge of these two functions completely characterized the solution.

To any real, Lorentzian solution, $Y = Y$, $\tilde{Y} = \bar{Y}$ we can associate a complex, self-dual solution characterized by $Y = Y$ and $\tilde{Y} = 0$, which his leads to a tetrad of the form \cite{Easson:2023dbk}
\be 
\begin{aligned} \label{self-dual-tetrad}
	& \lb = V du + dv + VY d\tilde{\zeta},  \\
	& \nb = du + Y d \tilde{\zeta}, \\
	& \mathbf{m} = d \tilde{\zeta},\\
	& \tbm = Y dv + d\zeta. 
\end{aligned}
\ee
\noindent In particular, the Kerr metric corresponds the functions \cite{mcintosh1988single}
\be \label{Y-eq}
 Y = \frac{v-u-ia-\sqrt{(v-u-ia)^2+4\zeta\tilde\zeta}}{2\tilde\zeta}, \qquad \tilde{Y} = \bar{Y}. 
\ee
The self-dual analog of Kerr then corresponds to the same $Y$, but with $\tilde{Y} = 0$, which leads to
\be \label{V-eq}
	V = \frac{-2 \sqrt{2} m}{\sqrt{(v-u - ia)^2+4 \zeta \tilde{\zeta}}}.
\ee 
The self-dual analog of Kerr is then described by the line element

\be
    ds^2 = 2dudv-2d\zeta d\tilde\zeta - \frac{2\sqrt{2}m}{\sqrt{(v-u-ia)^2+4\zeta\tilde\zeta}}\bigg(du + \frac{v-u-ia-\sqrt{(v-u-ia)^2+4\zeta\tilde\zeta}}{2\tilde\zeta}d\tilde\zeta\bigg)^2. \label{Self-Dual-Element}
\ee

When all of the coordinates are taken to be real and independent and the angular momentum is analytically continued $a \to -ia$, the line element \eqref{Self-Dual-Element} is real and Kleinian.\footnote{Of course, there are real values of the coordinates where the line element \eqref{Self-Dual-Element} becomes complex. This is generically the case for vacuum Kerr-Schild solutions with Kleinian signature, since $Y$ and $\tilde{Y}$ are defined as the roots of  algebraic equations involving the coordinates. Here we simply restrict the range of the coordinates to the region where the metric is real.} At this point, the rotation parameter $a$
can be removed from the metric by performing the shift $u' = u - ia$ or $v' = v + ia$ (before analytically continuing to imaginary $a$). The resulting metric can be expressed in Cartesian via (the Kleinian analog of) \eqref{eq:coor} as

\be \label{self-dual-Kerr}
    ds^2 = dt^2-dx^2+dy^2-dz^2-\frac{2m}{r}\Big(dt-dz-\frac{r-z}{x-y}d(x-y)\Big)^{\!2}
\ee
where $r^2\equiv x^2-y^2+z^2 = \frac12 ( (v-u)^2 + 4\zeta\tilde{\zeta})$.

We see that the angular momentum is pure gauge in the self-dual analog of Kerr, so without loss of generality, the rotation parameter $a$ may be set to zero in all of our equations. The fact that the rotation parameter can be gauged away with a simple shift of either the $u$ or $v$ coordinate is reminiscent of the observation  that the self-dual Kerr-Taub-NUT metric is diffeomorphic to Taub-NUT instanton \cite{Crawley:2021auj}, but here the diffeomorphism is trivial, taking a character reminiscent of the Newman-Janis shift relating the Schwarzschild and Kerr metrics \cite{Newman:1965tw}.

\section{Comparing the solutions} \label{sec:applyCK}

\subsection{Self-dual analog of Kerr}

A null tetrad for the self-dual analog of Kerr is then determined completely by equations  \eqref{self-dual-tetrad}, \eqref{Y-eq}, and \eqref{V-eq}, from which the  left Weyl scalars can be found to be

\be 
\begin{aligned}
	\Psi_0 &= 0, \\
	\Psi_1 &= 0, \\
	\Psi_2 &= \frac{2m}{r^3},\\
	\Psi_3 &= \frac{3\sqrt{2}m\tilde{\zeta}}{r^4},\\
	\Psi_4 &= \frac{6m\tilde{\zeta}^2}{r^5},
\end{aligned}
\ee
while the right Weyl scalars vanish $\tilde{\Psi}_I = 0$, as they must for a self-dual solution.
Then applying the transformations \eqref{NullRot} and \eqref{NullBst}, all scalars except $\Psi_2$ may be set to zero by choosing
\be 
	B_2 = \pm \frac{\tilde{\zeta}}{\sqrt{2}r},~ C_1 = 0. \label{eqn:M2_order0}
\ee
\noindent Choosing the positive solution, we find
\be 
	\Psi_2 = \frac{2m}{r^3}.
\ee

Notice that in this case the chosen form of the Weyl scalars are not affected by transformations in equation \eqref{NullBst}. Similarly, transformations of the form \eqref{NullRot} with $B_2$ and $C_1$ as in equation \eqref{eqn:M2_order0} are not affected by varying the parameters $B_1$ and $C_2$. Thus, the linear isotropy group at zeroth order is four dimensional, $dim~H_0 = 4$. The number of functionally independent components is one, $t_1 = 1$. 

Continuing to first order, we compute $C_{abcd;e}$. We may fix the remaining parameters $B_1$ and $C_2$ by setting several components of $C_{abcd;e}$ to zero through the choice:
\be
\begin{aligned}
B_1 = \frac{u-v + \sqrt{(u-v)^2 + 4 \zeta \tilde{\zeta}}}{2 \tilde{\zeta}},\quad C_2 = \frac{\tilde{\zeta}}{ 2 \sqrt{2} m - \sqrt{(u-v)^2 + 4 \zeta \tilde{\zeta}} }. \label{eqn:M2_order1}
\end{aligned}
\ee

\noindent The non-zero components may then be written as two distinct algebraically dependent sets:
\be 
\begin{aligned}
	&C_{1212;1} = - C_{1213;4} = C_{1234;1} =  C_{1324;1} = C_{1334;4} = -C_{3434;1}, \\
	& \\
	&C_{1212;2} = - C_{1224;3}= -C_{1234;2} = C_{1324;2} =  - C_{2434;3} = C_{3412;2} = - C_{3434;2},
\end{aligned} \nonumber 
\ee 
\noindent with the algebraically independent components: 
\be
\begin{aligned}
	& C_{1212;1} = \frac{3\sqrt{2}m}{r^4}, \\
	& C_{1212;2} =  \frac{3\sqrt{2}(r-2m)m}{r^5}.
\end{aligned}
\ee

We note that the transformations of the form \eqref{NullBst} have not been used explicitly. However, for the chosen null directions, the parameter $A_1$ can be fixed. For example, by setting $C_{1212;1} = 1$. However, $A_2$ cannot be fixed as it never appears at first order. Hence the linear isotropy group at first order is 1-dimensional, $dim~H_1 = 1$. The number of functionally independent invariants is still one $t_1 = 1$. 

As an alternative choice for $A_1$ we may instead ask that $C_{1212;1} = - C_{1212;2}$. For this choice, we find the following equation for $A_1$:
\be
    2 A_1^2 m \sqrt{2} - A_1^2 \sqrt{2}r - \sqrt{2}r =0
\ee
\noindent with the solution
\be
    A_1 = \pm \frac{\sqrt{r(r-2m)}}{r-2m}.
\ee
\noindent it follows that the condition $C_{1212;1} = - C_{1212;2}$ is only applicable when $r \geq 2 m$.

The algorithm continues to the second iteration. The set of non-zero components of the second covariant derivative of the Weyl tensor, $C_{abcd;ef}$, is naturally larger, although there are few new functional expressions and the components can be divided into three distinct classes of algebraically dependent components. The representatives of these classes will be displayed in section \ref{sec:conclusions}, but we note that there are no new functionally independent components, so that $t_2 =1$. Finally all non-zero components of $C_{abcd;ef}$ that require a contraction with $\mb$ must be contracted with $\tbm$, this implies that any boost with parameter $A_2$ has no effect on the components and it follows that $dim~H_2= 1$. 

\subsection{The Taub-NUT instanton} 
It was shown in \cite{Crawley:2021auj} that the Kleinian self-dual Taub-NUT metric can be written in spherical-like coordinates as\footnote{In what follows we will consider  the negative of the metric \eqref{Taub-NUT-instanton}, so that the resulting Cartan invariants share the same sign as the self dual analog of Kerr.}

\be\label{Taub-NUT-instanton}
ds^2 = \frac{\tilde{r} - \tilde{m}}{\tilde{r}+\tilde{m}}\left(dt - 2\tilde{m}\cosh\theta\,d\phi\right)^2 + \frac{\tilde{r}+\tilde{m}}{\tilde{r}-\tilde{m}}d\tilde{r}^2 - \left(\tilde{r}^2 - \tilde{m}^2\right)\left(d\theta^2 + \sinh^2\theta\,d\phi^2\right),
\ee 
where $\tilde{r} \geq -\tilde{m}$, $\theta \geq 0$, and $\phi$ and $t$ satisfy the periodicity conditions $\phi \sim \phi + 2\pi$ and $t \sim t + 4\pi \tilde{m}$, and furthermore, that the diffeomorphism
\bea
    \theta &=& 2\mathrm{\ arctanh}\left(\tanh\left(\tfrac{\theta'}{2}\right)\sqrt{\tfrac{r' - \tilde{m} - a}{r' - \tilde{m} + a}}\right), \\ \tilde{r} &=& r' + a\cosh\theta',
\eea 
maps \eqref{Taub-NUT-instanton}, to the Kleinian self-dual Kerr-Taub-NUT metric. The metric \eqref{Taub-NUT-instanton} has singularities at $r = \pm\tilde{m}$. The singularity at $r = -\tilde{m}$ is a curvature singularity, which can be seen from inspecting the Kretschmann scalar
\be 
R_{\mu\nu\rho\sigma}R^{\mu\nu\rho\sigma} = \frac{96 \tilde{m}^2}{(\tilde{r} + \tilde{m})^6}.
\ee
The status of the singularity at $\tilde{r} = \tilde{m}$ is less clear. In \cite{Crawley:2021auj}, it was argued that this singularity is a coordinate artifact. However, we argue in section \ref{sec:conclusions}
 that this may in fact be a physical singularity that is not captured by Kretschmann scalar. 

An obvious choice of orthonormal frame for the metric \eqref{Taub-NUT-instanton} is 
\be
\begin{aligned}
	\eb^1 &= \sqrt{\frac{\tilde{r}-\tilde{m}}{\tilde{r}+\tilde{m}}} (dt - 2 \tilde{m} 					\cosh\theta \,d\phi), \\
	\eb^2 &= \sqrt{\frac{\tilde{r}+\tilde{m}}{\tilde{r}-\tilde{m}}} dr, \\
	\eb^3 &= \sqrt{\tilde{r}^2-\tilde{m}^2} d\theta, \\
	\eb^4 &=  \sqrt{\tilde{r}^2-\tilde{m}^2} \sinh\theta \,d\phi,
\end{aligned}
\ee
so that 
\be
	ds^2 = (\eb^1)^2 + (\eb^2)^2 - (\eb^3)^2 - (\eb^4)^2. \label{metric-frame}
\ee
\noindent We may then select a complex null basis as, 
\be
\begin{aligned}
	& \mb_1 = \frac{1}{\sqrt{2}}(\eb^1 + i \eb^2), \quad 			\bar{\mb}_2 = \frac{1}{\sqrt{2}}(\eb^1 - i \eb^2), \\
	& \mb_2 = \frac{1}{\sqrt{2}}(\eb^3 + i \eb^4), \quad 			\bar{\mb}_2 = -\frac{1}{\sqrt{2}}(\eb^3 - i\eb^4).
\end{aligned}
\ee

Applying the transformations in equation \eqref{NullRot} and \eqref{NullBst}, we find the following non-zero Weyl scalars satisfy the type {\bf D}  condition when $C_1 = B_2 = 0$ yielding
\be
	\Psi_0 = \Psi_4 = \Psi_1 = \Psi_3 = 0, 
\ee
\noindent and
\be
	\Psi_2 = \frac{\tilde{m}}{(\tilde{r}+\tilde{m})^3}.  
\ee

To determine the linear isotropy group, $H_0$, with the constraints on the above parameters, we see that $A_1, A_2, C_2$ and $B_1$ do not appear in these expressions. Thus, at zeroth order, the dimension of the linear isotropy group is four, i.e., $dim~H_0 = 4$. The number of functionally independent invariants is, $t_0 = 1$. 

At first order, null transformations \eqref{NullRot} with $C_2 = B_1 = 0$ affect the form of the covariant derivative of the Weyl tensor and the majority of these components can be set to zero by setting $C_1 = B_2 = 0$, giving the following non-zero components as a single class of algebraically dependent components:  
\be 
\begin{aligned}
	&C_{1212;1} = - C_{1213;4} = C_{1234;1} =  C_{1324;1} = C_{1334;4} = -C_{3434;1} =  \\
    &-C_{1212;2} =  C_{1224;3}= C_{1234;2} = -C_{1324;2} =   C_{2434;3} = -C_{3412;2} =  C_{3434;2},
\end{aligned} \nonumber 
\ee 

\noindent where the algebraically independent component is
\be
	C_{1212;1} =  3 \sqrt{2} \, \tilde{M} \sqrt{\frac{\tilde{R} - 2\tilde{M}}{\tilde{R}^9}}.
\ee
\noindent where we have defined
\be
    \tilde{R}  = \sqrt{2} (\tilde{r}+\tilde{m}), \quad \tilde{M} = \sqrt{2} \tilde{m}. \label{eqn:tildeRM}
\ee

 We note that the parameter $A_1$ in the spin rotations in \eqref{NullBstC} can be fixed to $\pi/2$ to ensure that all quantities are real-valued. However, $A_2$ does not appear at first order and this implies that the linear isotropy group at first order is one, i.e., $dim~H_1 = 1$. The number of functionally independent components remains one, $t_1 = 1$. 

The algorithm must continue to second order to formally achieve the stopping conditions, $dim~H_1 = dim~H_2 = 1$ and $t_1 = t_2 = 1$. As in the previous example, the components of $C_{abcd;ef}$ belong to three distinct equivalence classes of algebraically dependent components. The distinct representatives will be displayed in the following section in order to prove the equivalence with the self-dual analog of Kerr.

\section{Conclusions} \label{sec:conclusions}

In this note we have studied two Kleinian spacetimes that have been conjectured to be equivalent \cite{Easson:2023dbk}. Previous arguments for their equivalence have relied on scalar polynomial curvature invariants constructed from their respective curvature tensors. However, due to limited knowledge of polynomial scalar curvature invariants involving covariant derivatives of the curvature tensor, the question of equivalence could not be fully answered. Using the Cartan-Karlhede algorithm \cite{Brooks:2015lba}, we are now able to answer the question in the affirmative without determining an explicit diffeomorphism carrying one line element to the other. 

In the previous section, we presented the necessary steps to invariantly characterize the self-dual analog of Kerr and the self-dual Taub-NUT spacetimes in terms of Cartan invariants. We will now review the main classifying quantities for both solutions. The discrete sequences for the dimension of the linear istropy group and the number of functionally independent components of the Weyl tensor its covariant derivatives at each iteration of the algorithm  are identical: 
\be \label{invariants}
\begin{aligned}
& \{ t_q \} = \{ 1, 1, 1\}, \\
& \{ dim~H_q \} = \{ 4, 1, 1\}.
\end{aligned}
\ee
\noindent In particular, using the formula in \eqref{IsoDim}, we find that both solutions admit 4 isometries. Finally, comparing the canonical forms of the curvature tensor, both solutions are of Petrov type {\bf D}, and so long as $r \geq 2 \sqrt{2} m$ in the self-dual analog of Kerr, we are able to choose the same canonical forms for the covariant derivatives of the curvature tensor, up to order 2. Interestingly, this inequality on the values of $r$ is not necessary when comparing with the equivalence of the double Kerr-Schild form of the self-dual Taub-NUT metric obtained by setting $\gamma = \lambda = 0$, $\epsilon = -1$ and $m = \ell$ in equation (32) of \cite{chong2005separability}. To be precise, it is always possible to globally choose $C_{1212;1}=1$ for both the self-dual analog of Kerr and the double Kerr-Schild form of the self-dual Taub-NUT solution in \cite{chong2005separability}, instead of $C_{1212;1}=-C_{1212;2}$ and that with this choice the resulting Cartan invariants agree at all points in the manifolds, hence the two metrics are globally equivalent. 

In contrast, the self-dual analog of Kerr and the form of the Taub-NUT instanton given by \eqref{Taub-NUT-instanton} agree only outside the horizons at $r=2m$ or $\tilde{r} = m$. To this point, we note that the determinant of the metric \eqref{Taub-NUT-instanton} vanishes at $\tilde{r} = \tilde{m}$, while the determinant of \eqref{self-dual-Kerr} is a non-zero constant. This is reminiscent of the situation one encounters when comparing a Schwarzschild black hole in Kerr-Schild form, (or even in standard Schwarzschild coordinates), with the form one obtains in isotropic coordinates. In the Kerr-Schild form, the metric determinant is constant (and in the standard Schwarzschild form, it vanishes only at the origin) whereas in isotropic coordinates, the metric takes the form
\be \label{isotropic}
ds^2_{\textrm{Isotropic}} = -\Big(\tfrac{\one-\tfrac{\EM}{\two\AR}}{\one +\tfrac{\EM}{\two\AR}}\Big)^{\!2}dt^2 + \left(\one + \tfrac{\EM}{\two\AR}\right)^4(d\AR^2 + \AR^2d\Omega^2).
\ee 
The determinant of \eqref{isotropic} vanishes on the horizon $R = M/2$. In fact, it is known that \eqref{isotropic} is only a faithful description of the \emph{exterior} of a Schwarzschild black hole, with the region $0 < R < M/2$ being diffeomorphic to the region $R >M/2$. Hence, the global structure of the line element \eqref{isotropic} is that of an Einstein-Rosen bridge connecting two asymptotically flat regions, rather than a black hole with a singularity at $R = 0$ \cite{Poplawski:2009aa, Gomes:2013fca, Gomes:2013bbl}. Indeed, while the Kretchmann scalar of \eqref{isotropic} is finite everywhere, it's curvature scalar is a delta-function with support on the horizon $R = M/2$ \cite{Poplawski:2009aa}, and hence, is not smooth at the horizon. We expect that a similar breakdown in the smooth structure of the solution \eqref{Taub-NUT-instanton} occurs at the horizon $\tilde{r} = \tilde{m}$, leading to the inequivalence of \eqref{Taub-NUT-instanton} and \eqref{self-dual-Kerr} within their horizons.

At this point, it must be noted that the agreement of the sequences of functionally independent components and linear isotropy groups summarized in \eqref{invariants} are not sufficient to conclude that \eqref{self-dual-Kerr} and \eqref{Taub-NUT-instanton} are (locally) equivalent---we must also compare the Cartan invariants and their classifying functions. In what follows, we will only examine the algebraically independent Cartan invariants. The location of the distinct Cartan invariants in the ordered set of components of the curvature tensor, and its covariant derivatives will be identical for both solutions.

Writing $\sqrt{2} r= ((u-v)^2+4\zeta \tilde{\zeta})^{\frac12} $, for $r \geq 2 m$, the unique Cartan invariants up to second order of the self-dual analog of Kerr may be written as:
\be
\begin{aligned}
& 0^{th} \text{ order: }  \frac{2m}{r^3}, \\
& 1^{st} \text{ order: } -3 \sqrt{2} m \sqrt{\frac{ r-2 m}{r^9}}, \\
&  2^{nd} \text{ order: } \frac{12 m}{r^5} - \frac{24 m^2}{r^6}, \quad \frac{30 m^2}{R^6}- \frac{12 m}{r^5}, \frac{18m^2}{r^6}-\frac{12 m}{r^5}.
\end{aligned} 
\ee

\noindent Whereas the the unique Cartan invariants up to second order of the self-dual Taub-NUT spacetime are:
\be
\begin{aligned}
& 0^{th} \text{ order: }  \frac{2\tilde{M}}{\tilde{R}^3}, \\
& 1^{st} \text{ order: } -3 \sqrt{2} \tilde{M} \sqrt{\frac{ \tilde{R}-2 \tilde{M}}{\tilde{R}^9}}, \\
&  2^{nd} \text{ order: } \frac{12 \tilde{M}}{\tilde{R}^5} - \frac{24 \tilde{M}^2}{\tilde{R}^6}, \quad \frac{30 \tilde{M}^2}{\tilde{R}^6}- \frac{12 \tilde{M}}{\tilde{R}^5}, \frac{18\tilde{M}^2}{\tilde{R}^6}-\frac{12 \tilde{M}}{\tilde{R}^5}
\end{aligned} 
\ee
\noindent with $\tilde{R}$ and $\tilde{M}$ defined in equation \eqref{eqn:tildeRM}. 

Due to the simple nature of the functionally independent Cartan invariant and the subsequent classifying functions, we see that formally both solutions have identical Cartan invariants. This is sufficient to conclude that the subset of the self-dual analog of Kerr where $ r \geq 2 m $ is equivalent to the self-dual Taub-NUT solution. Furthermore, a part of the diffeomorphism can be determined by equating the respective Cartan invariants in each spacetime to determine a relationship between the coordinates that appear in the invariants. The remaining coordinate transformations in the diffeomorphism could be determined by comparing the Killing vector-fields associated with the isometries. However, this is not the aim of the current note.

To the authors' knowledge, the use of the complex conjugate basis, $(\mb^1, \bar{\mb}^1,\mb^2, \bar{\mb}^2)$ has not been employed previously in the Cartan-Karlhede algorithm to classify Kleinian geometries. If one were to use the real-valued frame basis, $(\lb, \nb, \mb, \tbm)$, in the algorithm one would conclude that the two solutions were inequivalent. It is possible this could provide deeper insights into the canonical forms used in the alignment classification \cite{hervik2010pseudo}. More generally, we expect the Cartan-Karlhede algorithm to be useful for classifying Kleinian and Kerr-Schild geometries as examples of these structures proliferate in the mathematics and physics literature.

 \begin{acknowledgments}
We are grateful to Damien Easson, Tucker Manton, Michael Graesser, and Sigbjørn Hervik for thoughtful discussions. DM was supported by the Norwegian Financial Mechanism 2014-2021 (project registration number 2019/34/H/ST1/00636).
 \end{acknowledgments}

\bibliographystyle{JHEP}
\bibliography{bibliography}

\end{document}